\documentclass[multphys,vecphys]{svmult}

\usepackage{makeidx}     
\usepackage{graphicx}    
\usepackage{multicol}    

\makeindex             

\begin{document}

\title*{The Carbon content in the Galactic CygnusX/DR21
star forming region}
\author{H.~Jakob\inst{1},
R.~Simon\inst{1},
C.~Kramer\inst{1},
B.~Mookerjea\inst{1},
N.~Schneider\inst{2},
S.~Bontemps\inst{2} \and
J.~Stutzki\inst{1}}
\authorrunning{H. Jakob et al.}
\institute{
KOSMA, I.Physikalisches Institut, Universitaet zu Koeln, Germany
\and  Observatoire de Bordeaux, France}
%
%
\maketitle

\section{Introduction}

Observations of Carbon bearing species
are among the most important diagnostic probes of ongoing star formation. CO is
a surrogate for H$_2$ and is found in the vicinity of star formation sites. There,
[CI] emission is thought to outline the dense molecular cores and extend into the
lower density regions, where the impinging interstellar UV radiation field plays a critical
role for the dissociation and ionization processes. Emission of ionized carbon
([CII]) is found to be even more extended than [CI] and is linking up with the ionized medium.
These different tracers emphasize the importance of multi-wavelength studies to draw a
coherent picture of the processes driving and driven by high mass star formation.
Until now, large scale surveys
were only done with low resolution,
such as the COBE full sky survey, or were biased to a few selected
bright sources (e.g. Yamamoto et al. 2001, Schneider et al. 2003).
A broader basis of unbiased, high-resolution observations of [CI], CO, and [CII] may play a key role to probe the
material processed by UV radiation.


We here present a 2 deg$^2$ large-scale map of $^{13}$CO 2-1 (see Figure 1)
in the Cygnus X region and follow-up ($12'\times14'$) maps of the DR21 region in the two [CI] fine
structure lines and the $^{12}$CO 7-6 rotational transition.
Additional KOSMA observations of $^{12}$CO 6-5, $^{13}$CO 6-5, and $^{12}$CO 3-2 are
included for comparison.
We compare parts of the KOSMA data with a [CII] map taken with
KAO in 1994, $^{12}$CO 9-8, and $^{13}$CO 9-8 observations (Boreiko \& Betz 1991, also KAO) and
with publicly available ISO-LWS spectra at 7 distinct positions in the vicinity of DR21.

\section{The Observations: CI vs. CO}

The DR21 HII region/molecular cloud core has long been subject of detailed
studies, which have led to a comprehensive view of the region. Measuring several
parsecs in diameter at a distance of $1.5$ kpc (Bontemps et al., in prep.),
it harbors probably one of the most extended bipolar
outflow in the Galaxy.  The region contains a compact HII region at the
center and two extended lobes of shocked H$_2$ emission running along an axis
from NE to SW. The molecular material found in the lobes was presumably
ripped off a dense molecular ridge that is aligned N-S, and the wind
originating from DR21 is about to break through the cavity (Lane et al.
1990).

\vspace{0.5cm}

\begin{minipage}[b]{0.7\textwidth}
\includegraphics[height=8cm,angle=-90]{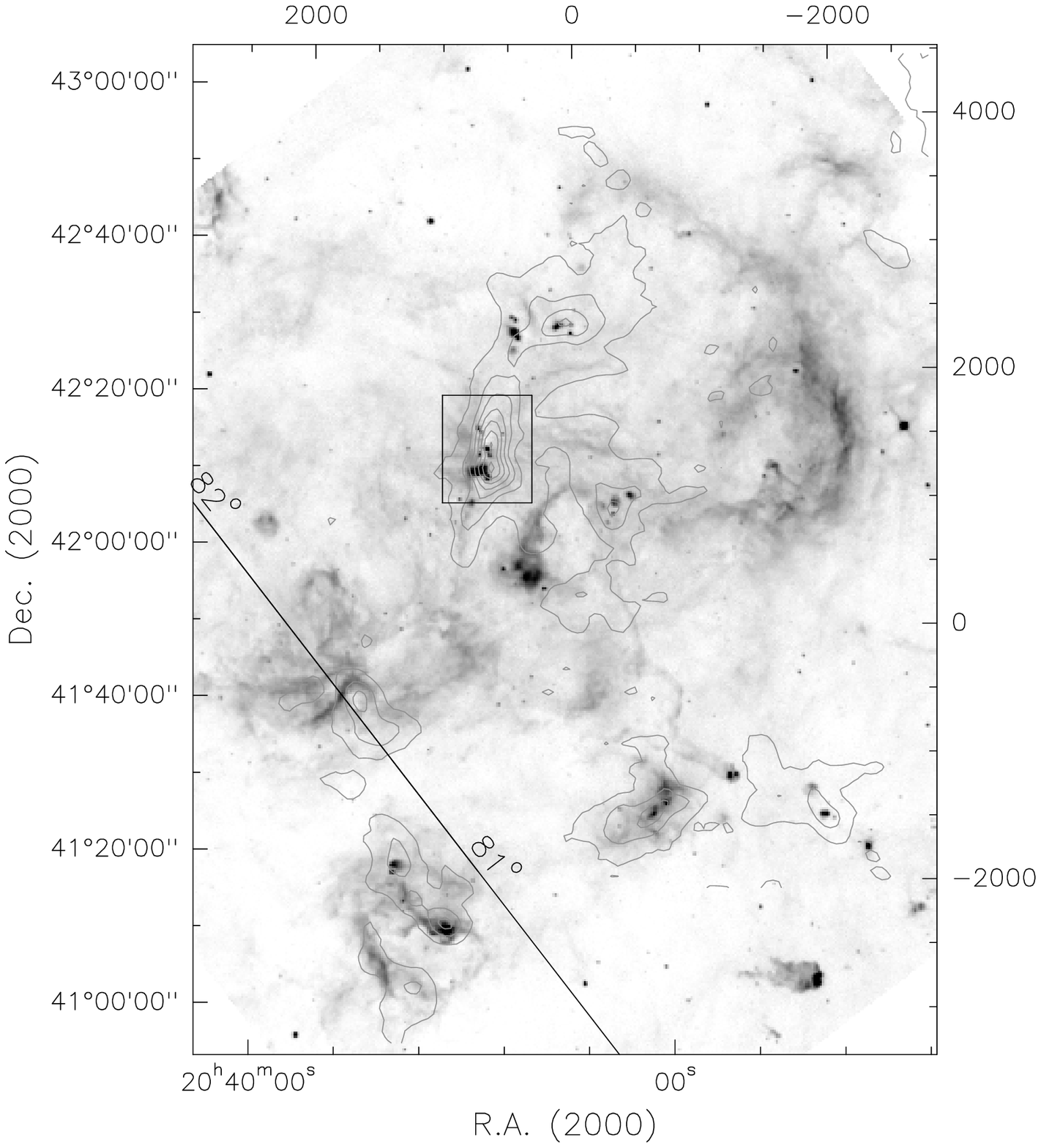}
\end{minipage}
\begin{minipage}[t]{0.24\textwidth}
{\bf Figure\,\,1:} KOSMA $^{13}$CO 2-1 velocity integrated intensity between $-6$ and 1 kms$^{-1}$
(contours) on top of an 8 $\mu$m MSX image.
The DR21 region mapped in both CI lines and 
CO 7-6 is marked by a box (cf. Figure 2).
\end{minipage}

\vspace{0.5cm}


Besides $^{13}$CO 2-1 (Figure 1), we have gathered observational data of both [CI] lines that are
assumed to be optically thin. We find, that the ratio of the integrated
intensity is almost constant at about 0.9 in the mapped region (compare with Figure 2).
This indicates that the [CI] lines have their origin in a warm environment. The emission
roughly follows the shape of the low-J $^{13}$CO maps. This clearly is in
contrast to the CO 7-6 map, which shows strong emission toward the DR21 core
and the outflow lobes, but not along the ridge in the vicinity of DR21(OH) and FIR1
(cf. Figure 3).
 
\vspace{0.5cm}

\begin{minipage}[b]{0.65\textwidth}
\includegraphics[height=3.5cm,angle=-90]{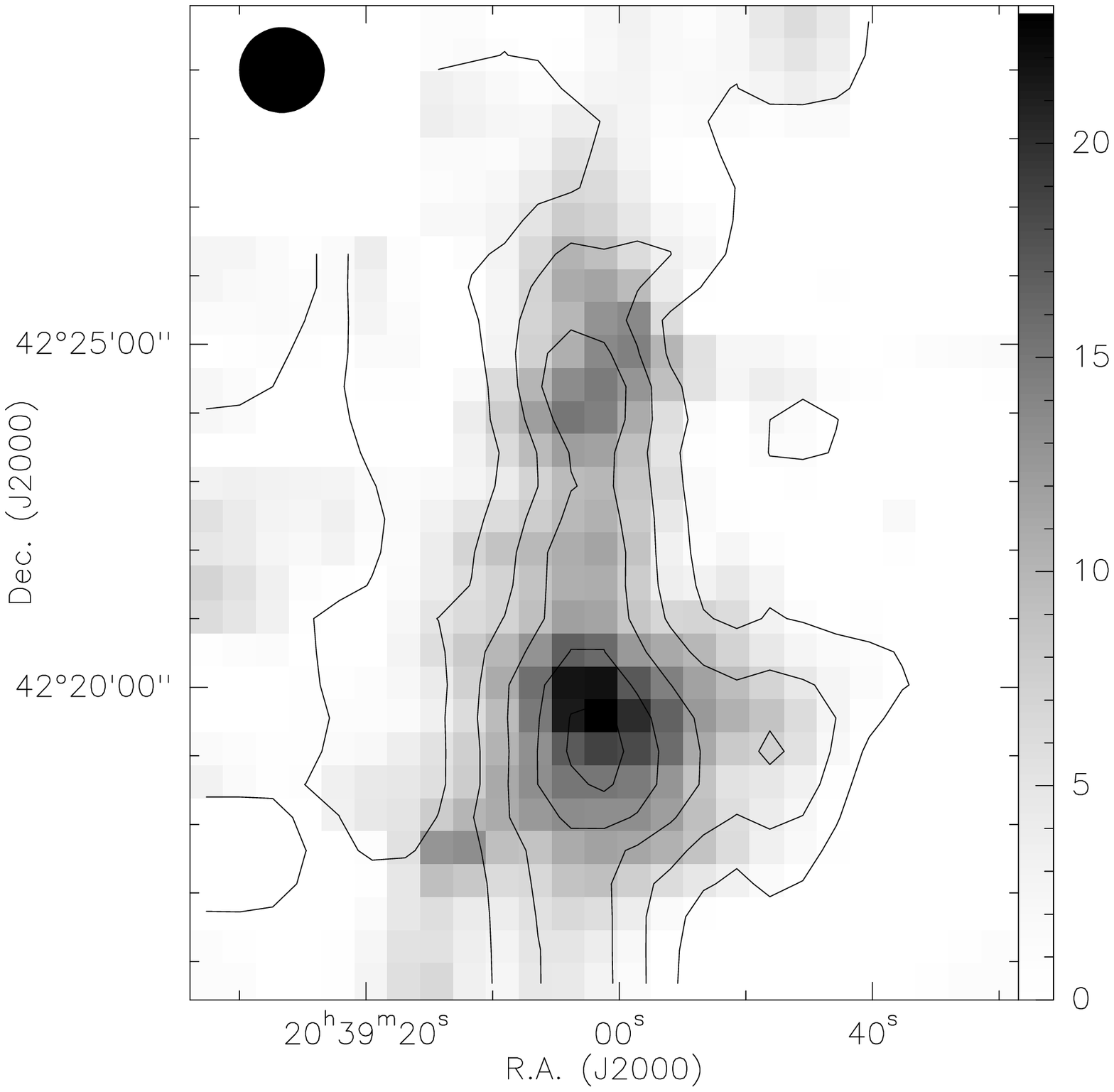}\nobreak
\includegraphics[height=3.5cm,angle=-90]{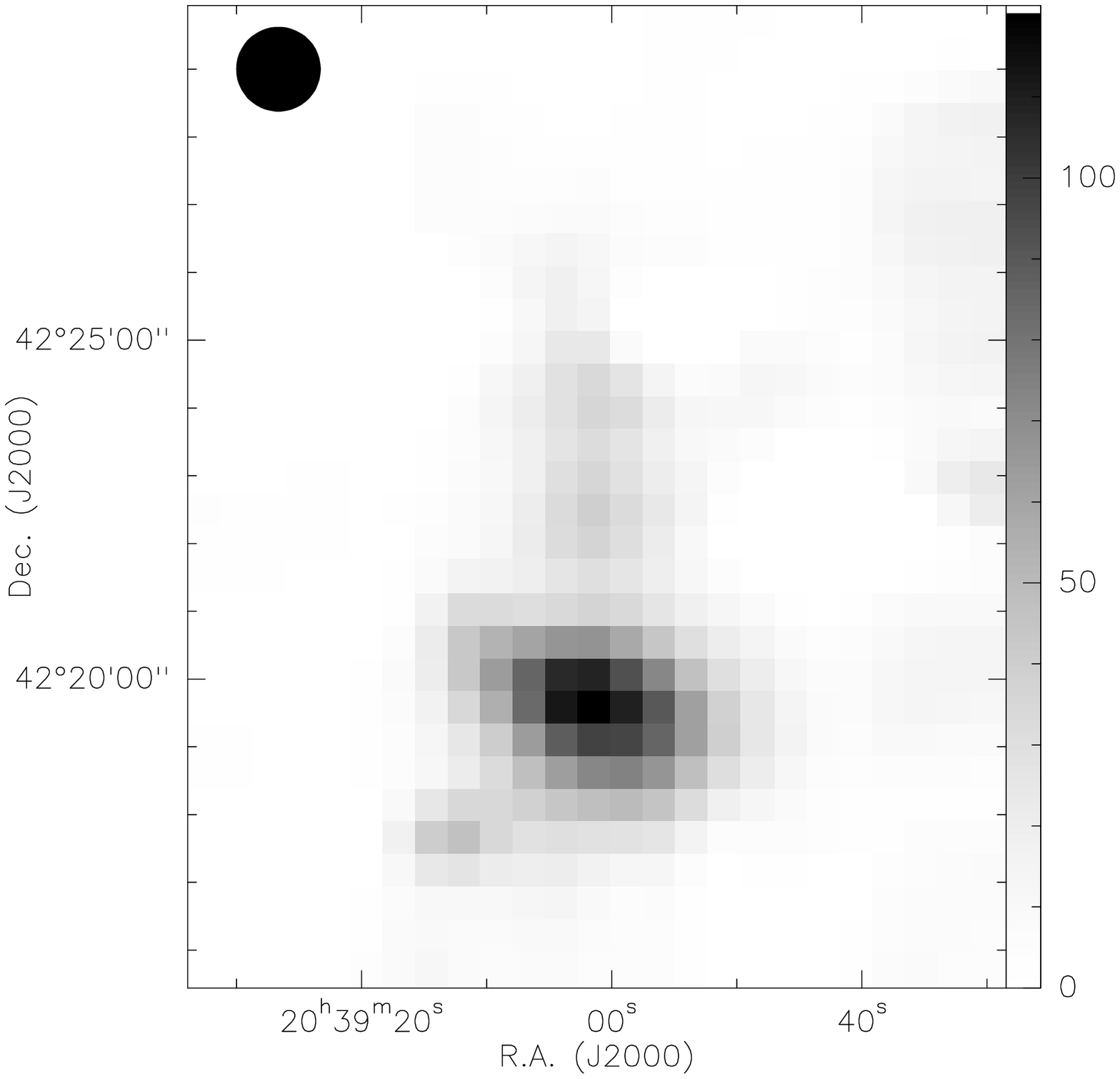}
\end{minipage}
\begin{minipage}[t]{0.3\textwidth}
{\bf Figure\,\,2:} KOSMA observations of [CI] $^3$P$_1-^3$P$_0$ (left, in contours),
[CI] $^3$P$_1-^3$P$_0$ (greyscale)
and
CO 7-6 (right) of the DR21/DR21(OH) region.
\end{minipage}

\section{The ISO LWS sample}
Seven ISO-LWS positions (see Figure 3) cover the core (\#2), the east and west lobe of the DR21 outflow
(\#1 and \#3) and two positions in between (DR21 East and West). The two nearby
sources DR21(OH) and DR21 FIR1 lie inside the dust lane to the north.

\vspace{0.4cm}
\begin{minipage}[t]{\textwidth}
\includegraphics[totalheight=11cm,angle=-90]{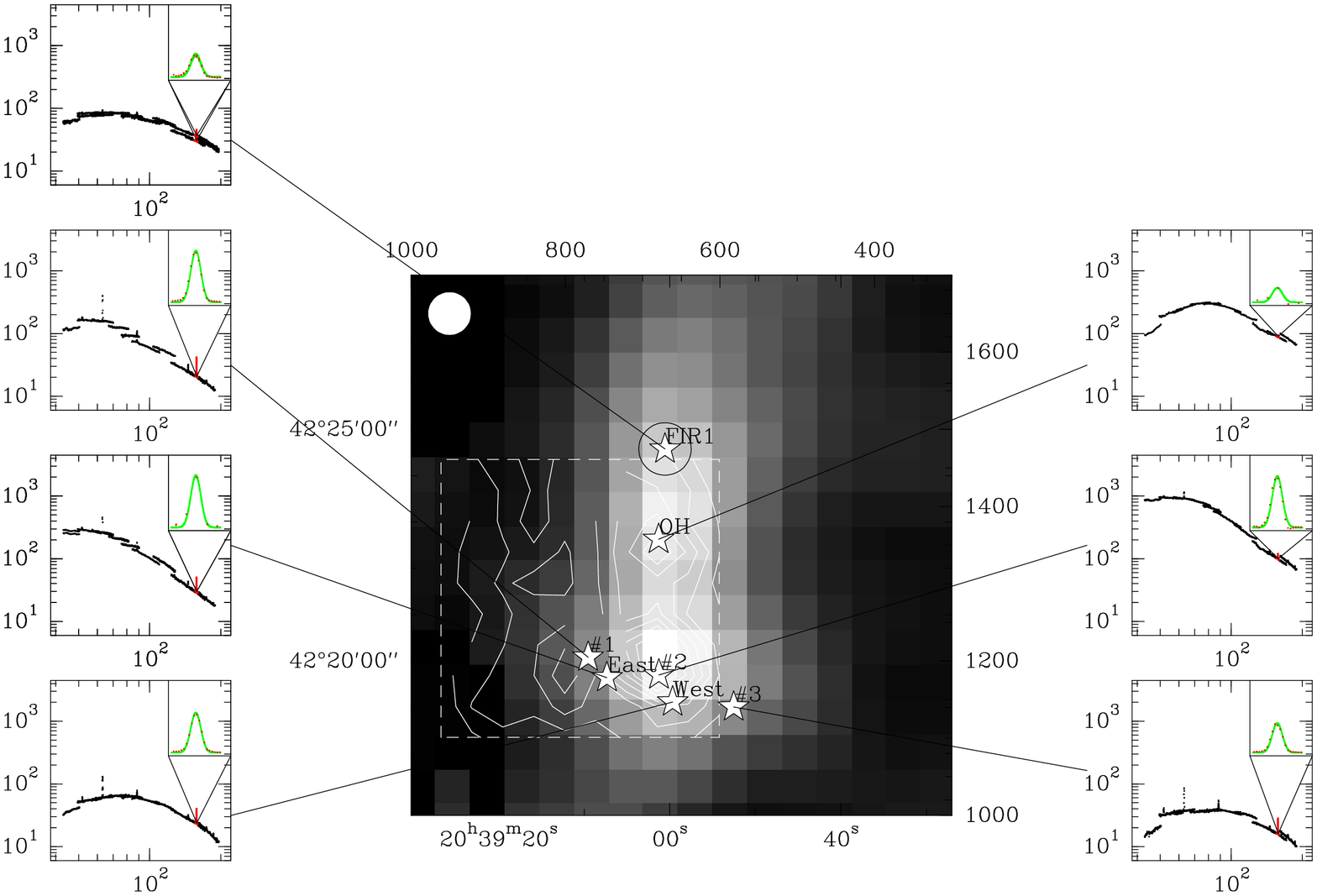}
\end{minipage}

\begin{minipage}[b]{0.95\textwidth}
{\bf Figure\,\,3:} Overlay of KOSMA $^{13}$CO 3-2 (gray) and KAO [CII] data (in contours). The
boxes show the ISO/LWS-FIR continuum and lines (e.g. high-J CO, [CII], [OI] and [OIII]) with a closeup
view on the (unresolved) [CII] line at 158$\mu$m on the right. Circles indicate the KOSMA and ISO beamwidths.
\end{minipage}

\vspace{0.5cm}

At position \#2, the continuum peak corresponds to L$_{44-197\mathrm{\mu m}}=4\times10^4$~L$_\odot$.
This result is consistent with earlier results (Harvey et al. 1977), when using the old distance estimate of 3 kpc instead of 1.5 kpc.
All major cooling lines are most
pronounced, but the line cooling efficiency (L$_\mathrm{Lines}$/L$_\mathrm{Cont}$) is much lower
than towards the eastern and western lobes (up to 1\% at position \#1). Especially to the east, 
the 63$\mu$m [OI] line intensity shows a strong east-west asymmetry and extends beyond
the continuum source.

\section{Molecular Gas: Physical Conditions}
We compare the observed $^{12}$CO and $^{13}$CO line intensities for position \#2 with a set of escape
probability models at different CO column and H$_2$ volume densities, as well as
kinetic temperatures. Two models are found to match the KOSMA and ISO
line intensities (see Figure 4). However, n(H$_2$) varies by a factor of 10
between the two solutions, whereas N(CO) and T$_\mathrm{kin}$ are better constrained. We
therefore included a $^{12}$CO and $^{13}$CO 9-8 observation of comparable beam width (Boreiko \& Betz 1991). It is
compatible with the $10^4$ cm$^{-3}$ model, but since $^{12}$CO 9-8 is not fitting at
all, we presume a calibration error for both lines.  
More valuable $^{13}$CO observations with high-J are planed with KOSMA.

\begin{minipage}[b]{\textwidth}
\includegraphics[height=11cm,angle=-90]{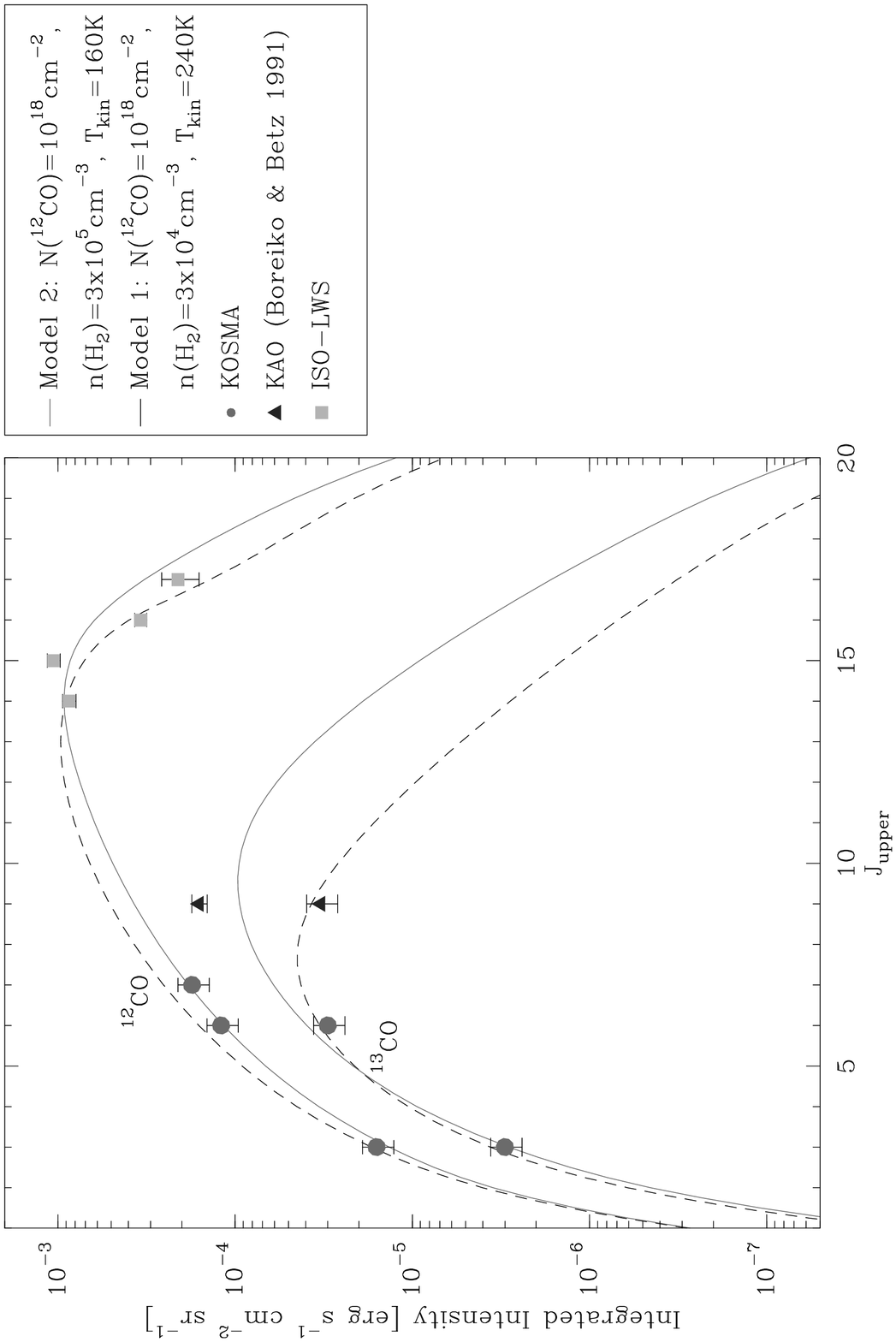}
\end{minipage}\nobreak

\begin{minipage}[t]{0.95\textwidth}
{\bf Figure\,\,4:} Observed CO line intensities (at position \#2) as a function of J$_\mathrm{upper}$. The two
lines represent best-fitting escape probability models assuming a $^{12}$CO/$^{13}$CO
abundance of 33.
The strong self absorption dip seen in the low-J and mid-J CO lines has not been accounted for.
\end{minipage}\nobreak

\section{Perspectives}

With this study, we aim at extending our observations of the emission due to
the major coolants of PDRs in Galactic star forming regions (e.g. S106
Schneider et al. 2003; W3 Kramer et al., in prep.). These data constitute
the basis for the application of radiative transfer and chemical PDR models
in order to 1. constrain the physical conditions in regions of high mass
star formation and 2. show whether all important physical processes are
taken into account in the current models to reproduce the observed line
intensities and ratios. Also, it allows to derive a consistent picture of
the photon dominated regions in the vicinity of massive star formation. This
serves as a preparatory work for future missions like SOFIA and Herschel,
which will provide a wealth of new information including velocity resolved
[CII] data.
\end{document}